\documentclass[12pt]{iopart}

\usepackage{iopams}  
\usepackage[dvips]{graphicx}

\begin{document}

\title[F-theory compactifications and central charges of BPS-states]{F-theory compactifications and central charges of BPS-states}

\author{Tetiana V. Obikhod }

\address{Institute for Nuclear Research NAS of Ukraine, 03680 Kiev, Ukraine}
\ead{obikhod@kinr.kiev.ua}
\vspace{10pt}
\begin{indented}
\item[] January 2016
\end{indented}

\begin{abstract}
F-theory, as Theory of Everything is compactified on Calabi-Yau threefolds or fourfolds. Using toric approximation of Batyrev and mirror symmetry of Calabi-Yau manifolds it is possible to present Calabi-Yau in the form of dual integer polyhedra. With the help of Gelfand, Zelevinsky, Kapranov algorithm were calculated the numbers of BPS-states in F-theory, and by application of Tate algorithm were determined the enhanced symmetries. As the result, any integral dual polyhedron representing a Calabi-Yau manifold, is characterized by its own set of topological invariants - the numbers of BPS states, whose central charges are classified by enhanced symmetries.
\end{abstract}
\maketitle
\normalsize

%
%
%
%
%

\section{Introduction}

F-theory or the "theory of everything" (Theory of everything, abbr. TOE) - hypothetical combined physical and mathematical theory that describes all known fundamental interactions. During the twentieth century,
It was proposed a lot of "theoryes of everything", but none of them could
go through experimental testing. The main problem of construction the scientific "theory of everything" is that quantum mechanics and general
Theory of Relativity (GTR) have different areas of their application.
Quantum mechanics is mainly used to describe the microcosm, and
general relativity applies to the macrocosm. Directly
the combination of quantum mechanics and special relativity in
single formalism (quantum relativistic field theory) leads to
divergence problem - the lack of final results for
experimentally testable variables. To solve this problem
was used the idea of renormalization. For some models
renormalization mechanism allows to build a very good working
theory, but the addition of gravity (ie the inclusion of the theory of general relativity as the limiting case of small fields and large distances) leads to divergences that still can not be removed. But it does not
mean that such a theory can not be constructed.

    Currently, the main candidate for a "theory of everything"
is an F-theory, which operates with a large number of measurements.
The impetus for this has become the Kaluza – Klein theory, which 
allows us to see that the addition of extra dimension to general theory 
of relativity leads to Maxwell's equations. 
Thanks to the ideas of Kaluza and Klein it was
possible to create theories that operates with  large dimensions.
Using of the extra dimensions proposed the answer to the question of
why the action of gravity is much weaker than
other types of interactions. The conventional answer is that
gravity exists in additional dimensions, so its effect as the
observable become weaker.

F-theory – string twelve-dimensional theory defined on
the energy scale of the order of  10 $^ {19}$ GeV  \cite{1.}. 
F-theory compactification  leads to a new type of vacuum, so for study SUSY
we must compactify F-theory on Calabi-Yau manifold.
Since there are a lot of Calabi-Yau manifolds, we are dealing with
a large number of new models implemented in low-energy
approximation. A singular manifold Calabi-Yau 
determines the physical characteristics of the topological solitonic
states that are interpreted as particles in high energy physics.
Essential for us is to present threefold Calabi-Yau
in the form of an elliptic fibration with singular layers,
that enables to use Kodaira’s classification of
singularities for elliptic bundles. To the type of singularities correspond
the sets of particles classified by enhanced symmetry, for which it is possible to find BPS states. Interpretation of these BPS states for the fiber bundles is presented in the paper.
The purpose of the article is the following. It is known that
Calabi-Yau manifolds can be represented as 
dual reflexive polyhedron with integer vertices.
To such manifold, on the one hand, you can associate a set of topological invariants - BPS states, calculated by application Gelfand, Zelevinsky, Kapranov algorithm, and on the other  hand - the enhanced symmetry obtained by applying Tate algorithm.
Thus, BPS states can be definitely characterized by a set of enhanced symmetries what is important to further searches for new physics at collider experiments in future.
As the singularities of  elliptic fibration are classified by enhanced groups and, at the same time, characterized by the number of BPS states,
which are determined by central charges, then with points on polyhedra of enhanced groups can be associated the central charge in analogy with the 
charge grid for electric and magnetic charges in Maxwell's electrodynamics.
   Let’s consider in more detail the compactification of F-theory 
to Calabi-Yau threefols.

\section{Compactification on Calabi-Yau threefolds and 
toric representation  of  threefolds }
Twelve-dimensional space, describing space-time and
internal degrees of freedom is represented as following:
\[R^6 \times X^6 \ ,\]
where $R^6$ - six-dimensional space-time, on which acts
conformal group $SO (4, 2)$ аnd $ X^6$ - compact threefold, 
three-dimensional complex manifold Calabi-Yau. 
 	Let's consider the weighted projective space defined as follows:
\[P^4_{\omega_1,\ldots,\omega_5 }=P^4/Z_{\omega_1}\times \ldots \times Z_{\omega_5}\ , \]
where $P^4$ - fourdimensional projective space,
$Z_{\omega_i}$ - cyclic group of order $\omega_i$.
On a weighted projective space $P^4_{\omega_1,\ldots,\omega_5 }$
is determined quasihomogeneous
polynomial $W(\varphi_1, \ldots, \varphi_5)$,
called superpotential, which satisfies the homogeneity condition
\[W(x^{\omega_1}\varphi_1, \ldots, x^{\omega_5}\varphi_5)=x^d W(\varphi_1, \ldots, \varphi_5)\ ,\]
where $d=\sum\limits_i\omega_i$, $\varphi_1, \ldots, \varphi_5 \in P^4_{\omega_1,\ldots,\omega_5 } $.
The set of points $p\in P^4_{\omega_1,\ldots,\omega_5 } $,
satisfying $W(p)=0$ forms Calabi-Yau threefold 
$X_d(\omega_1, \ldots, \omega_5)$ \cite{2.}.

\subsection{Toric manifolds as an extensions of weighted projective
  spaces }

The simplest examples of toric varieties are projective spaces 
$P^{2}$ and $P^{(2,3,1)}$, where $P^{2}$
is defined as follows
\[P^{2}=\frac{C^{3}{\backslash}{0}}{C{\backslash}{0}},\]
where division into $C{\backslash}{ 0}$ means 
the identification of points in complex space $C$, connected
by equivalence relation
\[(x,y,z)\sim (\lambda x, \lambda y, \lambda z )\]
\[\lambda \in  C{\backslash}{ 0},\]
$ x, y, z $ are called homogeneous coordinates. Elliptic curve in $P^{2}$ is
described by the Weierstrass equation
\[y^2z=x^3+axz^2+bz^3.\]
A similar description can be given for $P^{(2,3,1)}$, which in 
contrast to $P^{2}$ is represented by the following equivalence relation:
\[(x,y,z)\sim (\lambda^2 x, \lambda^3 y, \lambda z )\]
\[\lambda \in  C{\backslash}{ 0},\]
and Weierstrass equation has the form
\[y^2=x^3+axz^4+bz^6.\]
The elliptic Calabi-Yau manifold can be described by
Weierstrass form
\[y^2=x^3+xf(z)+g(z),\]
which describes an elliptical fibration (parameterized by
$(y, x)$) over the base, where $f(z), g(z)$ - functions on the basis 
\cite{1.}.
24 parameters on $P^1$ associated with the functions $ f (z), g (z) $ are 
specified by zeros of the discriminant.
Then in some divisors $ D_i $ the layer is degenerated. Such divisors are the zeros of the discriminant
\[\Delta=4f^3+27g^2.\]
Singularities of Calabi-Yau manifold - are singularities of its
elliptic fibration. These singularities are
encoded in the polynomials $f, g $ and their type determines the gauge group and matter content of the compactified F-theory. Classification of singularities of elliptic fibrations was given by Kodaira and presented in table \ref{jlab1}.
\begin{table}
\caption{\label{jlab1}Kodaira's classification of singularities of elliptic fibrations }
\begin{center}
\begin{tabular}{|c|c|c|}\hline
$ord(\Delta)$&Fiber type & Singularity type \\ \hline
0&smooth&no\\ \hline
n&$I_n$&$A_{n-1}$ \\ \hline
2&$II$&no \\ \hline
3&$III$&$A_{1}$ \\ \hline
4&$IV$&$A_{2}$ \\ \hline
n+6&$I^*_n$&$D_{n+4}$ \\ \hline
8&$IV^*$&$E_{6}$ \\ \hline
9&$III^*$&$E_{7}$ \\ \hline
10&$II^*$&$E_{8}$ \\ \hline
\end{tabular}\\
\end{center}
\end{table}
\normalsize

Classification  of  the fibers of an elliptic fibrations is presented in figure 1.
\begin{figure}[ht]
\begin{center}
\includegraphics[width=0.5\textwidth]{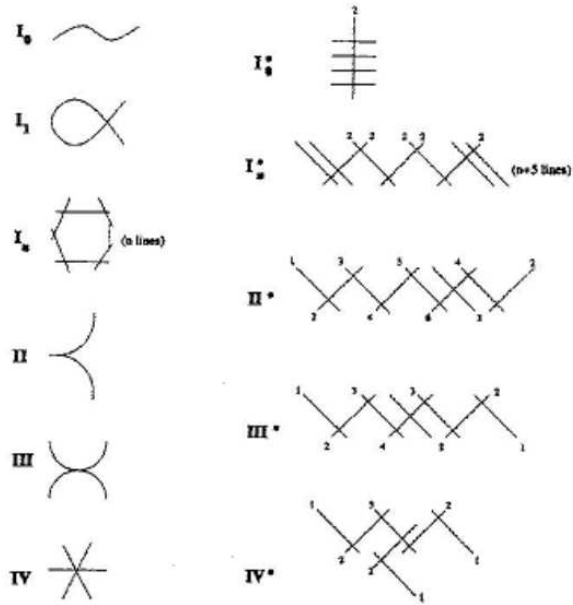}
\caption{Classification of elliptic fibers.}
\end{center}
\end{figure}

$P^{2}$ and $P^{(2,3,1)}$ may be represented by diagrams with vectors
$v_x, v_y, v_z$ in some lattice, such that
\[q_xv_x+q_yv_y+q_zv_z=0\ ,\]
where $q_x, q_y, q_z$ are exponents. 

Since the possible singular sets of Calabi-Yau manifold may be the points,
which are singularities of type $C^3/Z_{N_s}$ or
curves - singularities of type $C^2/Z_{N_s}$, 
both types of singularities and their
blow-up can be described by methods of toric geometry.
To describe toric variety $P^4_{\omega_1,\ldots,\omega_5 }$,
let’s consider integer polyhedron $\Delta\in R^n$. In this case, 
we can determine a simplicial reflexive polyhedron
\[\Delta(\vec{\omega}):=\left\{(x_1,\ldots, x_{n+1})\in R^{n+1}|
\sum\limits_{i=1}^{n+1}\omega_ix_i=0, x_i \geq -1 \right\}\]
Complex d-dimensional toric variety is defined by combinatorial data 
$\Delta$, called fan. A finite non-empty set $\Delta$, called a fan, 
is determined by a combination of convex 
rational polyhedral cones $\sigma$ in $R^{n+1}$
\[\sigma = R_{\geq}\vec{n_1}+\ldots + R_{\geq}\vec{n_r}\ .\]
If \\
1)	each face of a cone in $\Delta$ belongs to $\Delta$ and \\
2)	the intersection of any two cones in $\Delta$ is a face of each.

	Integer polyhedron $\Delta$ is called reflexive polyhedron
\cite{3.} if the corresponding dual polyhedron $\bigtriangledown$
\[\bigtriangledown=\left\{(y_1,\ldots , y_{n+1})|
\sum\limits_{i=1}^{n+1}x_iy_i\geq -1, \ (x_1,\ldots , x_{n+1})\in \Delta \right\}\]
is also integer. This property of polyhedra is connected with mirror symmetry of Calabi-Yau manifolds \cite {4.}. The vertices of a simplicial reflexive polyhedron $\Delta(\omega_i)$ are determined by the weights of $P^4(\omega_i)$,
since the degree $d$ of Calabi-Yau threefold $X_d(\omega_1, \ldots, \omega_5)$
satisfies the condition $d=\sum\limits_i\omega_i$. Examples of construction of reflexive polyhedra through the Calabi-Yau weights are given in \cite {5.}.

	Let’s consider the holomorphic three-form $\Omega(\psi)$ 
of threefold Calabi-Yau $X$ as a function of $\psi_i$ - coordinates on the complex Calabi-Yau space. Their derivatives are elements of group $H^3(X)$. After the integration of elements in $H^3(X)$, we get
linear differential equations for the periods $\Pi$, the 
Picard-Fuchs equations, which allows us to calculate the Yukawa couplings. 
In terms of Mori generators $l(\theta)$, satisfying
$\sum\limits_i l_i\omega_i=0$ and 
according to Gelfand, Kapranov and Zelevinsky algorithm \cite{6.},
thanks to mirror symmetry between Kahler and complex Calabi-Yau manifols,
we can write  Picard-Fuchs equation with periods $\Pi(x)$ as \cite{4.}:
\begin{equation*}
\hspace*{-3cm}
\left\{\Pi_{l_j^{(k)}> 0}\Biggl(\Pi_{i=0}^{l_j^{(k)}-1}(\theta_j-i)\Biggr)-
\Pi_{i=1}^{|l_0^{(k)}|}(i-|l_0^{(k)}|-\theta_0)\Pi_{l_j^{(k)}<0, j\ne 0}
\Biggl(\Pi_{i=0}^{|l_j^{(k)}|-1}(\theta_j+l_j^{(k)}-i)\Biggr)x_k\right\}
\tilde{\Pi}(x)=0\ ,
\end{equation*}
where $\theta_i=x_id/dx_i$, $x_i$ - are algebraic coordinates
on the moduli space of the complex structure of Calabi-Yau manifold.

	The principal parts of the Picard-Fuchs operators could have, in particular,
the form \cite {7.}:
\begin{eqnarray}
L_1&=&3\theta_1^2-\theta_1\theta_2+\theta_2^2,\nonumber \\
L_2&=&\theta_2^2, \nonumber \\
L_3&=&\theta_3^2,\nonumber \\
L_4&=&\theta_2^2+4\theta_2\theta_3+4\theta_3^2-3\theta_2\theta_4-6\theta_3\theta_4+9\theta_4^2\ .\nonumber
\end{eqnarray}
Yukawa couplings are:
\[K_{\tilde{t_i}\tilde{t_j}\tilde{t_k}}(\tilde{t})=
\frac{1}{\omega_0(x(\tilde{t}))^2}\sum\limits_{l,m,m}
\frac{\partial x_l}{\partial \tilde{t_i}}
\frac{\partial x_m}{\partial \tilde{t_j}}
\frac{\partial x_n}{\partial \tilde{t_k}}K_{x_lx_mx_n}(x(\tilde{t}))\]
and could be overwritten by a variable $q_i=e^{\tilde{t_i}}$:
\[K_{\tilde{t_i}\tilde{t_j}\tilde{t_k}}(\tilde{t})=K^0_{ijk}+
\sum\limits_{n_i}\frac{N({n_i})n_in_jn_k}{1-\prod\limits_lq_l^{n_l}}
\prod\limits_lq_l^{n_l}\ ,\]
where $\tilde {t_i}$ - Kahler space coordinates
$ x_i $ - coordinates of complex mirror manifold.
Here $n_i=\int\limits_Ch_i$ is an integer and not
necessarily positive, in particular, for singular varieties.
That is, the solution of the Picard-Fuchs equations makes
it possible to calculate the Yukawa coupling constants, which
are expressed in terms of the numbers  $n_i$. $n_1$
- is the number of rational curves of degree 1, $n_2$ - the number of rational curves of degree 2 etc. In general, $n_i $- numbers of BPS-states through which is determined the central charge and the mass of the solitonic objects. Thus, knowing Mori generators we can find the principal part of the Picard-Fuchs operators, through which are calculated numbers of BPS-states.

	In summary, it must be stressed that toric presentation of Calabi-Yau manifolds makes it possible to calculate the topological invariants - 
	BPS-states.

\section {Enhanced symmetry in F-theory}
F-theory allows geometrical and physical interpretation
of solitonic states in terms of geometrical singularities
and enhanced symmetry \cite{8.}.
Tate et al. proposed algorithm which allows
to extract the enhanced symmetry from toric description of
elliptic Calabi-Yau manifold. 
	This algorithm allows to read off the Dynkin
diagram from the dual polyhedron $\bigtriangledown$ 
that realizes toric description of elliptic Calabi-Yau manifold according
to toric  Batyrev’s approximation \cite{3.}.
Using the technique of \cite{8.},  dual polyhedron $^3\bigtriangledown$, representing Calabi-Yau is divided by 
triangle $^2\bigtriangledown$ on the top and bottom
\[\bigtriangledown=\bigtriangledown^H_{bot}\cup \bigtriangledown^{k=1}_{top}\ ,\]  
where $\bigtriangledown^H_{bot}$ depends on enhanced gauge group 
$H$ and $\bigtriangledown^{k=1}_{top}$ depends on $k$.
For fourfolds of type
\[X_{18k+18}(1,1,1,3k,6k+6,9k+9)\]
the gauge groups are written in the following way \cite{9.}:
\begin{center}
$ H \times SU(1) $ \hspace*{1cm} for \hspace*{1cm} 
$ k=1 $ ,\\
$ H \times SO(8) $ \hspace*{1cm} for \hspace*{1cm} 
$ k=2 $ , \\
$ H \times E_{6} $ \hspace*{1.8cm} for \hspace*{1cm} 
$ k=3 $ , \\
$ H \times E_{7} $ \hspace*{1.8cm} for \hspace*{1cm} 
$ k=4 $ , \\
$ H \times E_{8} $ \hspace*{1.8cm} for \hspace*{1cm} 
$ k=5 $ , \\
$ H \times E_{8} $ \hspace*{1.8cm} for \hspace*{1cm} 
$ k=6 $ . \\
\end{center}

	Thus, solitonic states, characterized by BPS-states, 
as singularities of Calabi-Yau manifolds may
be classified by enhanced symmetry as to each type of Calabi-Yau,
presented in the form of dual polyhedron,
corresponds its enhanced symmetry.

\section{Conclusion}
We have given the definition of Calabi-Yau hypersurfaces 
in weighted projective spaces through their weights
and presented Kodaira’s classification of singularities of elliptic fibrations. 
Application of Batyrev’s toric approach,
and  Gelfand, Kapranov, Zelevinsky algorithm 
made it possible to calculate the number of BPS states
characterizing the solitonic objects in the F-theory.
Consideration of Calabi-Yau using Tate’s algorithm 
enables to associate solitonic states with enhanced symmetries of F-theory. Thus, toric presentation of Calabi-Yau through Batyrev’s
toric approximation enables, on the one hand, to calculate BPS-states,
and on the other, to calculate the enhanced symmetry of the polyhedron,
describing massless solitonic states in F-theory.
The main result of the article is reduced to the conclusion that 
we get an adequate treatment of central charges 
of the BPS-states as elements on the polyhedron 
connected whith the enhanced symmetries.


\newpage
\section*{References}
\begin{thebibliography}{99}
\bibitem{1.} Vafa C., {\it Evidence for F-theory}, arXiv: hep-th/9602022; \\
Morrison D. R. and  Vafa C. 1996 Compactifications of F-theory on Calabi-Yau threefolds (I) {\it Nucl. Phys. B} 473 74-92; \\
 Morrison D. R. and  Vafa C. 1996 Compactifications of F-theory on Calabi-Yau threefolds (II) {\it Nucl. Phys. B} 476 437-69.
\bibitem{2.} Klemm A. and  Schimmrigk R. {\it Landau-Ginzburg string vacua}, hep-th/9204060.
\bibitem{3.}  Batyrev V. V. 1993 Variations of the Mixed Hodge Structure of Affine Hypersurfaces in Algebraic
Tori, {\it Duke Math. J.} 69 349-409.
\bibitem{4.}  Hosono S. ,  Klemm A. ,  Theisen S. and Yau S.-T. 1995  Mirror symmetry, mirror map and applications
to complete intersection Calabi-Yau spaces {\it Nucl. Phys. B} 433 501-54.
\bibitem{5.} Hosono S. ,  Klemm A. ,  Theisen S. and Yau S.-T. 1995 Mirror Symmetry, Mirror Map and Applications to Calabi-Yau Hypersurfaces
{\it Commun. Math. Phys.} 167 301-50.
\bibitem{6.} Gel’fand I. M., Zelevinsky A. V. and Kapranov M. M. 1990 
Generalized Euler integrals and A-hypergeometric functions {\it Adv. Math.} 84 255-71.
\bibitem{7.}  Malyuta Yu. and  Obikhod T. 2002 BPS-States in F-Theory {\it Ukr. Math. J.} 54 Issue 9 pp 1550-1555.
\bibitem{8.} Bershadsky M.,  Intriligator K.,  Kachru S.,  Morrison D. R.,  Sadov V. and Vafa  C., {\it Geometric Singularities and Enhanced Gauge Symmetries}, hep-th/9605200;\\
 Candelas P. and  Font A., {\it Duality Between the Webs of Heterotic and
Type II Vacua}, hep-th/9603170;\\
 Perevalov E. and  Skarke H., {\it Enhanced Gauge Symmetry in Type II
and F-Theory Compactifications: Dynkin Diagrams From Polyhedra},
hep-th/9704129;\\
 Katz S.,  Morrison D. R.,  Schäfer-Nameki S. and 
 Sully J., {\it Tate's algorithm and F-theory}, arXiv: 1106.3854 [hep-th].
\bibitem{9.} Malyuta Yu. and  Obikhod T., {\it Compactifications of F-Theory on Calabi-Yau Fourfolds}, arXiv:hep-th/9803241.
\end{thebibliography}
\end{document}